\documentclass[reprint,prl,aps]{revtex4-1}

\usepackage{graphicx}
\usepackage{amsmath,amssymb}
\usepackage{epstopdf}

\begin{document}

\title{The unsteady evolution of localized unidirectional deep water wave groups}
\author{Will Cousins}
\affiliation{Department of Mechanical Engineering, Massachusetts Institute of Technology, 77 Massachusetts Ave., Cambridge, MA, USA}
\author{Themistoklis P. Sapsis}
\thanks{Corresponding author; e-mail: sapsis@mit.edu; }
\affiliation{Department of Mechanical Engineering, Massachusetts Institute of Technology, 77 Massachusetts Ave., Cambridge, MA, USA}

\begin{abstract}
We study the evolution of localized wave groups in unidirectional water wave envelope equations (nonlinear Schrodinger (NLS) and modified NLS (MNLS)). These localizations of energy can lead to disastrous extreme responses (rogue waves). Previous studies have focused on the role of energy distribution in the frequency domain in the formation of extreme waves. We analytically quantify the role of spatial localization, introducing a novel technique to reduce the underlying PDE dynamics to a simple ODE for the wave packet amplitude. We use this reduced model to show how the scale-invariant symmetries of NLS break down when the additional terms in MNLS are included, inducing a critical scale for the occurrence of extreme waves.
\end{abstract}

\maketitle

Understanding extreme events is critical due to the catastrophic damage they inflict.  Important examples of extreme events are: freak ocean waves \cite{dysthe2008,muller2005}, optical rogue waves \cite{solli2007}, capsizing of ships \cite{kreuzer2006}, and extreme weather/climate events \cite{neelin2010,majda2010}.  In this work, we address the formation of freak/extreme waves on the surface of deep water.  These waves have caused considerable damage to ships, oil rigs, and human life \cite{haver2004,liu2007}.

Extreme waves are rare, and available data describing them is limited.  Thus, analytical understanding of the physics of their triggering mechanisms is critical.  One such mechanism is the Benjamin-Feir modulation instability of a plane wave to small sideband perturbations.  This instability, which has been demonstrated experimentally \cite{chabchoub2011}, generates huge coherent structures by soaking up energy from the nearby field \cite{benjaminFeir1967,zakharov1968,osborne2000}.  The ocean surface, however, is much more irregular than a simple plane wave.  The Benjamin-Feir \emph{Index} (BFI), the ratio of surface amplitude to spectral width, measures the strength of the modulation instability in such irregular fields.  For spectra with large BFI, nonlinear interactions dominate, resulting in more extreme waves than Gaussian statistics would suggest.  However, a large BFI does not provide precise spatiotemporal locations where extreme events might occur.

In large BFI regimes, spatially localized wave groups of modest amplitude focus, creating the extreme waves.  Thus, to complement the frequency-based approaches described above, it is essential to understand the precise role of \emph{spatial} localization in extreme wave formation.  In addition to providing insight into the triggering mechanisms for extreme waves, this analysis will allow the development of new spatiotemporal predictive schemes.  Specifically, by understanding which wave groups are likely to trigger an extreme wave, one could identify when and where an extreme wave is likely to occur, in a manner similar to that of Cousins and Sapsis \cite{cousinsSapsis2014} for the MMT model \cite{majda1997}. By analyzing the evolution of spatially localized fields, the authors found a particular length scale that was highly sensitive for the formation of extreme events.  By measuring energy localized at this critical scale, the authors reliably predicted extreme events for meager computational expense.

Two commonly used equations to model the envelope of a modulated carrier wave on deep water are the Nonlinear Schrodinger Equation (NLS) and the modified NLS equation (MNLS) \cite{dysthe1979}.  The focusing of localized groups is well understood for NLS (see work of Adcock et. al. \cite{adcock2009,adcock2012} and Onorato et. al. \cite{onorato2003}).  In this Letter, we study the less understood wave group focusing properties in the MNLS model. That is, given a wave group of a particular amplitude and length scale, we determine if this group will focus and lead to an extreme wave.

We find two striking differences between NLS and MNLS dynamics. First, due to a lack of scale invariance in MNLS, there is a minimal focusing length scale where wave groups below this scale do not focus.  Second, the higher order nonlinear terms of MNLS equation greatly inhibit focusing for some large amplitude groups.  That is, there is a considerably smaller set of wave groups that would lead to an extreme event in MNLS in comparison to NLS.  These features are critical for understanding realistic extreme waves, as MNLS is significantly more accurate in reproducing experimental results when compared with NLS \cite{goullet2011,lo1985}.  

We explain this difference in NLS and MNLS focusing analytically by using a single mode, adaptive projection where the length scale of the mode is allowed to vary with time.  We close this model by enforcing conservation of $\mathcal{L}^2$ norm, which follows from the envelope equations.  This drastically simplifies the relatively complex MNLS PDE, yielding a single ODE for the group amplitude. This reduced model agrees favorably with direct numerical simulations.  Furthermore, the simplicity of this ODE model allows us to analytically explain various aspects of group evolution in MNLS, such as the existence of a minimal focusing length scale and the smaller family of focusing groups relative to NLS.
%

NLS \cite{zakharov1968} describes the evolution of the envelope of a slowly modulated carrier wave on the surface of deep water:
\begin{align}
	\frac{\partial u}{\partial t} + \frac{1}{2} \frac{\partial u}{\partial x} + \frac{i}{8} \frac{\partial^2 u}{\partial x^2} + \frac{i}{2} |u|^2 u = 0
	\label{eq:NLS}
\end{align}
where $u$ is the wave envelope, $x$ is space, and $t$ is time. Dysthe developed the modified NLS equation (MNLS) by incorporating higher order terms \cite{dysthe1979}:
\begin{align}
  \begin{split}
  \frac{\partial u}{\partial t}& +\frac{1}{2}\frac{\partial u}{\partial x} + \frac{i}{8} \frac{\partial^2 u}{\partial x^2} - \frac{1}{16} \frac{\partial^3 u}{\partial x^3} + \frac{i}{2} |u|^2 u \\&+ \frac{3}{2} |u|^2 \frac{\partial u}{\partial x} + \frac{1}{4} u^2 \frac{\partial u^*}{\partial x} + i u \frac{\partial \phi}{\partial x} \Big|_{z=0} = 0
	\end{split}
	\label{eq:MNLS}
\end{align}
where $\phi$ is the velocity potential, which may be expressed explicitly in terms of $u$ by solving Laplace's equation \cite{trulsen1996}. To study the evolution of spatially localized groups, we use initial data of the form $u(x,0) = A_0 \text{sech}(x/L_0)$.

\begin{figure}
	\centering
	\includegraphics[width=0.85\columnwidth]{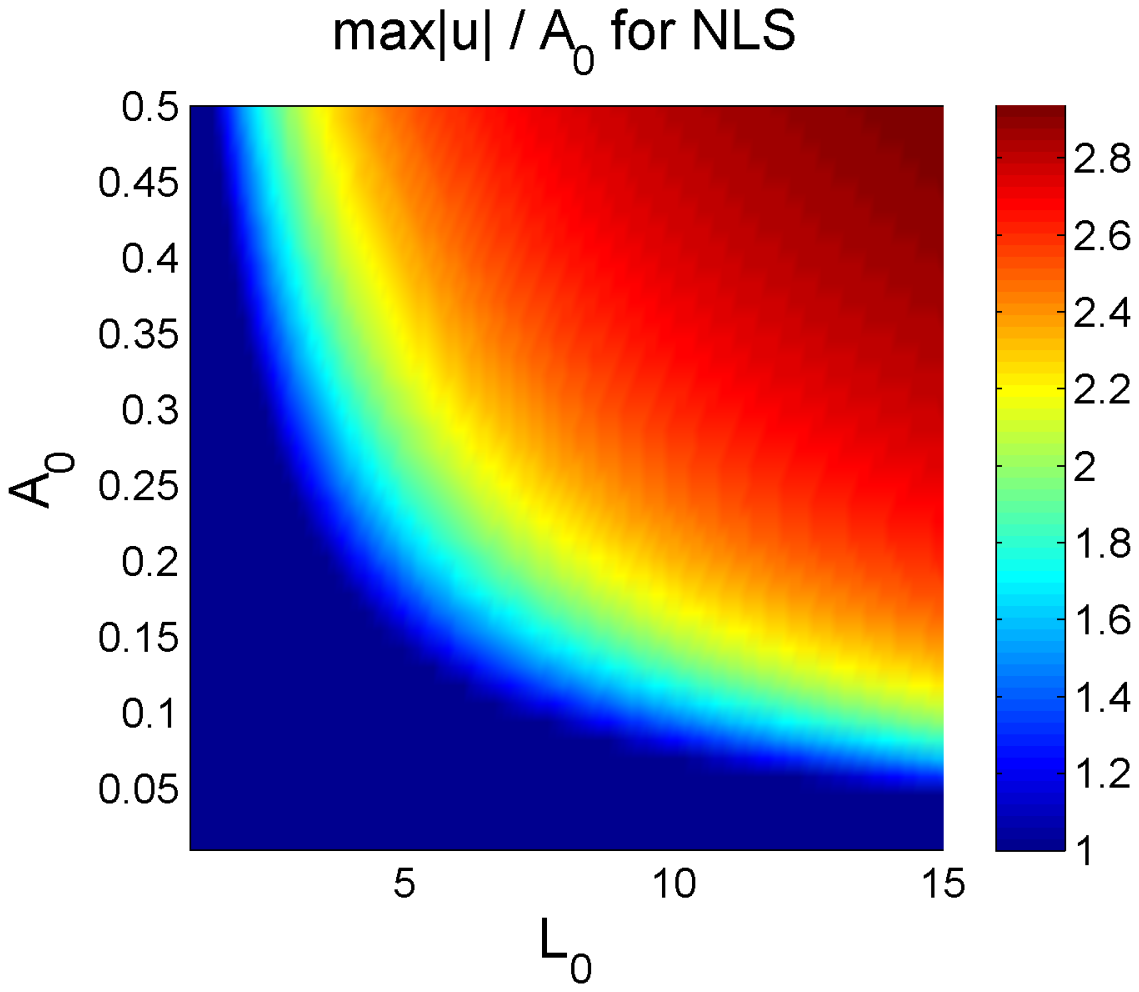} \\
	\includegraphics[width=\columnwidth]{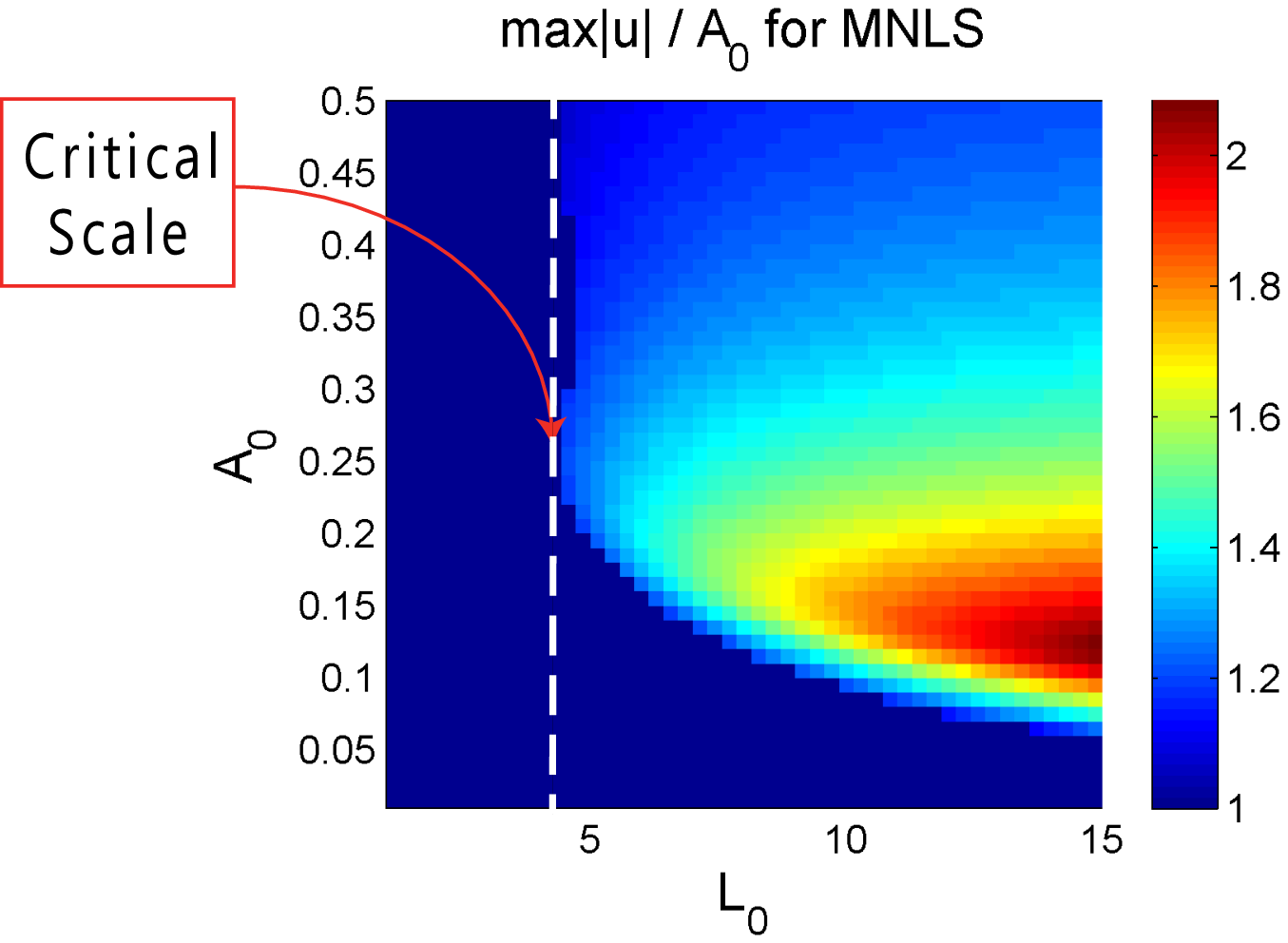}
	\caption{Ratio of first spatiotemporal local maximum amplitude divided by initial amplitude for NLS (top) and MNLS (bottom).}
	\label{fig:relGrowth}
\end{figure}

Using this family of initial conditions, we numerically compute the value of the first spatiotemporal local maximum of $|u|$.  In Figure~\ref{fig:relGrowth}, we display the value of this local maximum amplitude divided by the initial amplitude $A_0$. This quantity is 1 when defocusing occurs and the amplitude decreases.  Values of this ratio larger than 1 indicate that the associated group focuses, increasing in amplitude.  We mention that the top portion of our Figure~\ref{fig:relGrowth} is similar to Figure 1 of Onorato et. al. \cite{onorato2003}, where a similar topic was studied for the NLS equation.

For NLS, for each length scale there is an exact soliton solution with $A_0 = 1/(\sqrt{2}L_0)$, where the wave group shape is constant in time. If the initial amplitude is smaller than this solitonic amplitude, then the group broadens and its amplitude decreases.  If the initial amplitude is larger than this solitonic level, then the group focuses and increases in amplitude.  This behavior is qualitatively the same for all length scales due to the scale invariance of NLS.  Furthermore, the degree of focusing increases for larger amplitudes.  That is, for all $L_0$, $A_{max}/A_0$ is an increasing function of $A_0$.  

For MNLS, the situation is more complex (Figure~\ref{fig:relGrowth}, bottom). Similar to NLS, for some $(A_0,L_0)$ we do have an appreciable degree of focusing--thus MNLS posesses a mechanism for generating extreme waves.  However, this behavior is \emph{not} qualitatively the same for all length scales due to the additional nonlinear terms that lead to the breaking of scale invariance.  In particular, there is a minimum focusing length scale, where groups narrower than this length scale do not focus, regardless of how large their initial amplitude may be.  Morever, even when the length scale is larger, $A_{max}/A_0$ is not a monotonically increasing function of $A_0$.  There is thus a finite range of amplitudes that lead to significant focusing.  Furthermore, certain groups that do focus do so in a weaker sense compared to NLS. We mention that some amplitudes in Figure~\ref{fig:relGrowth} exceed the well known physical maximum wave steepness of $\approx0.4$ \cite{toffoli2010}.  However, NLS and MNLS dynamics do show substantial differences for much lower, physically relevant amplitudes.

To develop an approximate model for NLS we approximate solutions as $u(x,t) = A(t) \text{sech} \left((x-\frac{1}{2}t)/L(t)\right)$, which move at speed $1/2$ as this is the group velocity for the NLS equation.  Applying the ansatz for $u$ and projecting the equation to estimate $d |A|^2 /dt$ results in a trivial equation:
\begin{align*}
	\frac{d |A|^2}{dt} = A^* \frac{dA}{dt} + A \frac{dA^*}{dt} = 0.
\end{align*}

\begin{figure}
	\centering
	\includegraphics[width=\columnwidth]{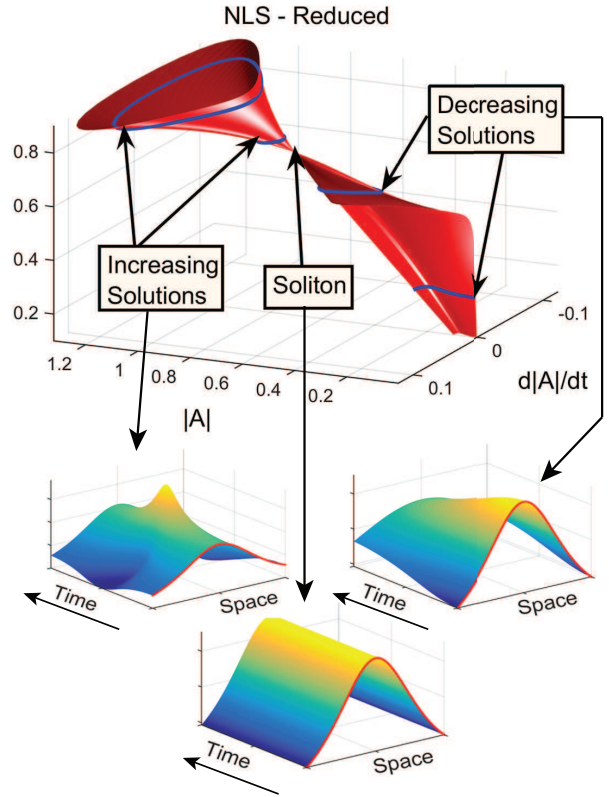}
	\caption{Top: surface described by solutions to the reduced order model (\ref{eq:NLSProj}) with various initial amplitudes, $L_0=1$.  Bottom:  predicted evolution for various wave groups.}
	\label{fig:phaseSurf_NLSProj}
\end{figure}

This equation is not helpful (although it is correct--we do observe the initial growth rate of groups to be zero in the full NLS).  We differentiate the NLS equation (\ref{eq:NLS}) to obtain the second time derivative of $|u|^2$.  We apply the ansatz for $u$, multiply the equation by the hyperbolic secant and integrate over the real line.  This is not sufficient to close the system as we have allowed both amplitude and length scale to vary with time. NLS conserves the integrated squared modulus of $u$, which implies $L(t) = L_0 |A_0 / A(t)|^2$.  Using this dynamical constraint, we obtain the following equation for $A(t)$:
\begin{align}
  \frac{d^2|A|^2}{dt^2} = \frac{K}{|A|^2} \left( \frac{d|A|^2}{dt} \right)^2 + \frac{3|A|^2 (2|A|^2L^2-1)}{64L^2}
	\label{eq:NLSProj}
\end{align}
where $K=(3\pi^2-16)/8$.  We are interested in trajectories of (\ref{eq:NLSProj}) with initial conditions $|A(0)|^2 = A_0^2$ and $d|A|^2/dt|_{t=0}=0$. Our reduced model has the correct, solitonic, fixed point $A_0 = 1 / (\sqrt{2}L_0)$, and predicts focusing if the initial amplitude is larger than this solitonic value, and decay if the initial amplitude is smaller (consistently with \cite{adcock2009,adcock2012}).
The family of solutions to (\ref{eq:NLSProj}) describes a surface in the coordinates $(A_0,|A|,d|A|/dt)$.  We plot this ``phase surface'' in Figure~\ref{fig:phaseSurf_NLSProj}.  The solitonic fixed point is clearly visible.  Solutions with $A_0$ larger than the solitonic value grow to a new maximum and oscillate periodically between this new maximum and $A_0$.  Interestingly, values of $A_0$ just less than the solitonic amplitude decrease initially but oscillate periodically in time, never exceeding the initial amplitude (Adcock et. al. made similar observations \cite{adcock2009}). A comparison with a direct simulation of NLS reveals similar behavior, although for NLS we have energy leakage away from the main group.

Applying this methodology to MNLS is more complicated as there is an amplitude dependent group velocity.  To address this, we express the solution as $u(x,t) = A(t) \text{sech}((x-ct)/L(t))$, where $c$ is an unknown constant group velocity.  We first find the second temporal derivative of $|u|^2$ in a coordinate frame moving with group velocity $c$. Projecting the resulting equation gives
\begin{align}
  	\begin{split}
  	&\frac{d^2|A|^2}{dt^2} = \frac{K}{|A|^2} \left( \frac{d|A|^2}{dt} \right)^2 - \\
	&c^2\frac{3|A|^2}{2L^2} -c\frac{3|A|^2(-7|A|^2L^2-4L^2-1)}{8 L^4} \\
	&- \frac{3|A|^2}{L^6}(980|A|^4L^4+832|A|^2L^4+392|A|^2L^2 \\
	&+256L^4+160L^2+43)
  	\end{split}
	\label{eq:MNLSProjWithVelocity}
\end{align}

where again $L(t) = L_0 |A_0 / A(t)|^2$. We must also prescribe a value for $c$. Consider the envelope $|u|$ of a generic wave packet, where, for each time $t$ there is a single local maximum whose spatial position is given by $x=bt$.  If we consider the evolution of $|u|$ along the ray $x'=ct$, we have that $|u(x'=ct,t)| < |u(x=bt,t)|$ if $c\neq b$. Thus, for wave group $|u|$ traveling with speed $b$, the growth of $|u|$ along the ray $x=ct$ is maximized (with respect to $c$) when we set $c=b$.  To select $c$ in the reduced order model (\ref{eq:MNLSProjWithVelocity}), we apply this criteria, choosing the value of $c$ that maximizes the right hand side of (\ref{eq:MNLSProjWithVelocity}), which governs the growth of $|A|$. This gives
\begin{align}
	c = \frac{7A_0^2L_0^2 + 4L_0^2+1}{8L_0^2}.
	\label{eq:MNLSGroupVelocity}
\end{align}
As expected, $c$ tends to the NLS group velocity of 1/2 as $L_0$ becomes large. Substituting (\ref{eq:MNLSGroupVelocity}) for $c$ in (\ref{eq:MNLSProjWithVelocity}) gives the following equation for $|A|^2$:
\begin{align}
  	\begin{split}
  	&\frac{d^2|A|^2}{dt^2} = \frac{K}{|A|^2} \left( \frac{d|A|^2}{dt} \right)^2 - \frac{3|A|^2}{2048L^6}(196|A|^4L^4\\&-64|A|^2L^4+168|A|^2L^2+32L^2+27).
        \end{split}
	\label{eq:MNLSProj}
\end{align}

\begin{figure}
	\centering
	\includegraphics[width=0.9\columnwidth]{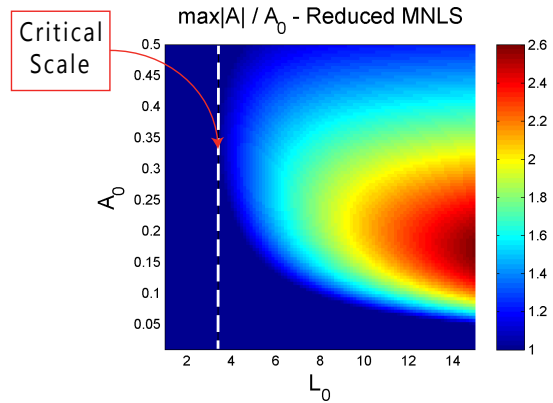}
	\caption{Maximal value of $A(t)$ relative to $A_0$ for solutions of the reduced order MNLS model (\ref{eq:MNLSProj}).  This compares favorably direct numerical simulations of MNLS (Figure~\ref{fig:relGrowth}, bottom).}
	\label{fig:relGrowth_MNLSProj}
\end{figure}

The reduced order MNLS equation (\ref{eq:MNLSProj}) has a bifurcation at $L = \sqrt{14/4} + \sqrt{35/16}\approx 3.35$.  If $L_0 < L^*$, the right hand side of (\ref{eq:MNLSProj}) will initially be negative, regardless of how large $A_0$ may be.  Thus, groups with length scale less than $L^*$ do not grow. This is precisely the behavior we observe in numerical simulations of the full MNLS equation (Figure~\ref{fig:relGrowth}).  To illustrate these analytical results, we solve the reduced equation numerically for various $L_0$ and $A_0$ and display $A_{max}/A_0$ in Figure~\ref{fig:relGrowth_MNLSProj}. 

For $L_0>L^*$, (\ref{eq:MNLSProj}) has two fixed points.  The amplitude will grow only when $A_0$ is \emph{between} these two fixed points.  Even for some focusing groups, the degree of focusing will be limited--the presence of the larger fixed point limits the amplitude growth relative to NLS, agreeing with direct numerical simulations of the MNLS PDE.  Each of these two fixed points suggests an envelope soliton of MNLS.  Although numerical simulations suggest that the lower amplitude fixed point does correspond to a soliton, the larger amplitude fixed point does not and is thus an artifact of the reduced order model.

\begin{figure}
	\centering
	\includegraphics[width=\columnwidth]{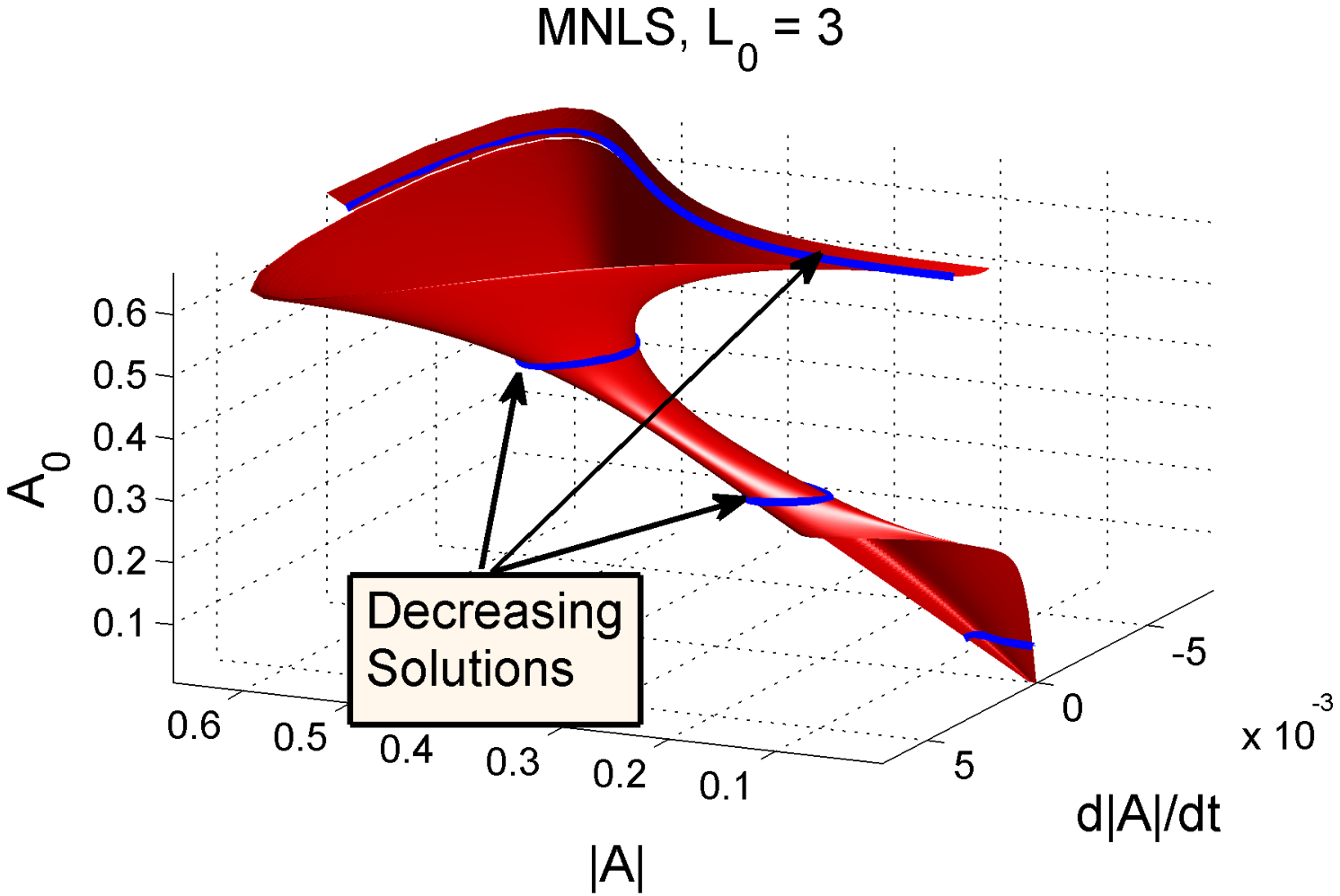} \\ [1em]
	\includegraphics[width=\columnwidth]{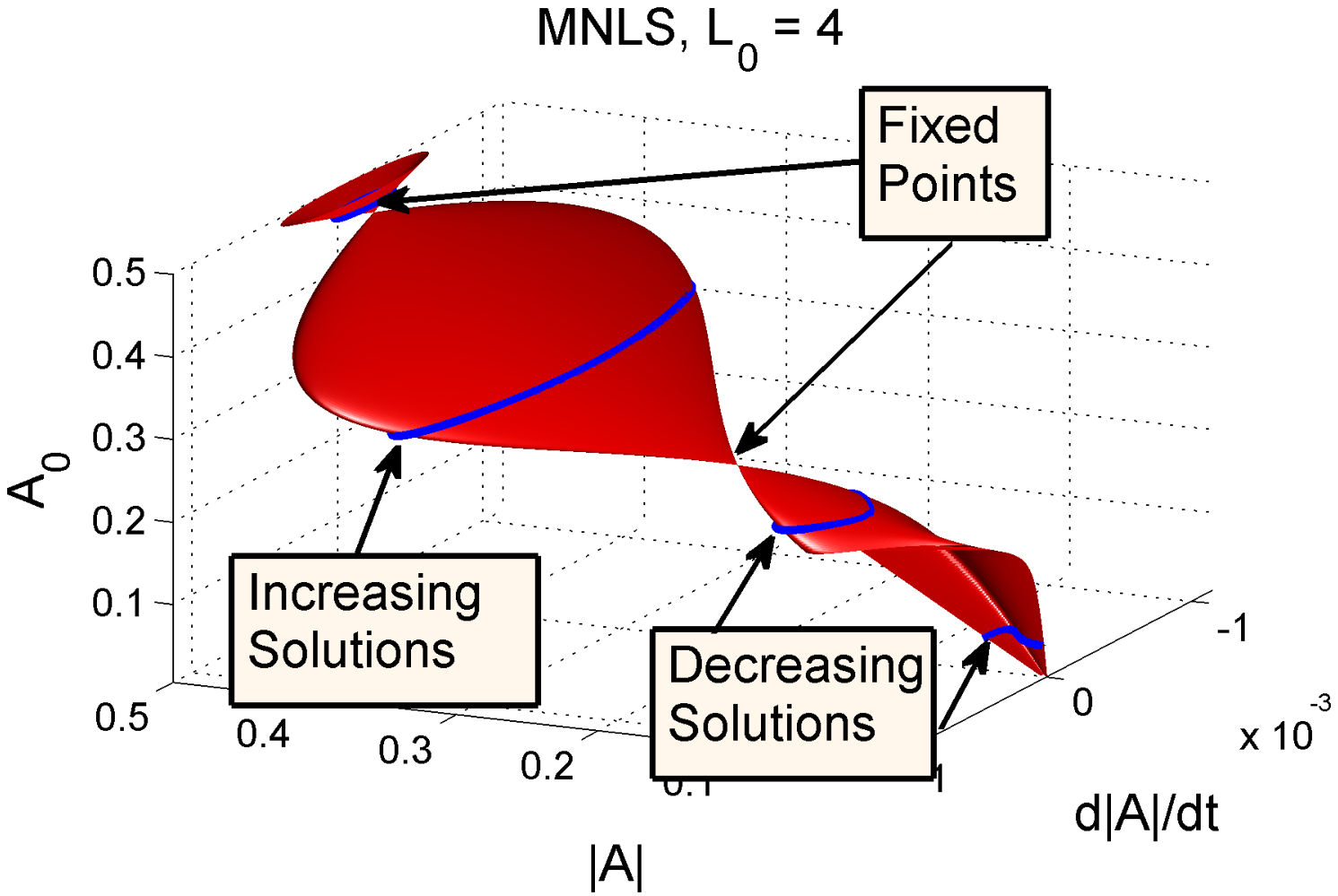}
	\caption{Phase surface diagrams for solutions to (\ref{eq:MNLSProj}) with $L_0 = 3$ (top) and $L_0 = 4$ (bottom). The bottom figure shows the emergence of two fixed points.}
	\label{fig:phaseSurf_MNLSProj}
\end{figure}

To illustrate the dynamics for MNLS, we display phase surfaces in Figure~\ref{fig:phaseSurf_MNLSProj} for $L_0 = 3$ and $L_0=4$.  For $L_0=3 < L^*$, all groups decay, with some groups oscillating periodically and others decaying monotonically.  For $L_0=4>L^*$, two fixed points emerge.  Between these fixed points is the region of focusing, with groups in this region increasing in amplitude, but with smaller increase compared with NLS. 

Considering the surface elevation, the wave crest height amplification can be \emph{larger} in MNLS due to higher-order terms in the formula for reconstructing the elevation from the envelope \cite{slunyaev2013}.  Crest-to-trough wave height amplification is lower in MNLS, agreeing with our observations of the dynamics of the envelope $|u|$ in this Letter.  The relevance of the crest height vs crest-to-trough height depends on the particular application of interest.

In summary, we developed a new approach for the analytical understanding of one-dimensional wave group evolution. Using a single-mode adaptive projection, we derived a simple ODE that mimics the dynamics of the underlying PDE remarkably well. The key of our approach allowing the localized mode to adaptively adjust its length scale to respect the conservative properties of the PDE. The order model explains a number of salient, scale-varying features of group evolution in MNLS. 

Compared with existing methods, our approach provides a large amount of information while being simple to implement.  For comparison, the BFI is simple to compute but does not provide the rich information of our approach.  Methods based on the Inverse Scattering Transform (IST) provide complete information but are complicated to implement \cite{slunyaev2006,osborne2005,islas2005}.  Additionally, the IST is not applicable to two-dimensional wave dynamics where the governing equations are not integrable \cite{osborne2010Book}.  The approach presented here is similar in spirit to existing soliton perturbation approaches for NLS \cite{fabrikant1980,kath1995}.  However, these approaches either consider only small perturbations about soliton solutions or require theoretical machinery unavailable for MNLS.

The main limitation of our approach is the assumption of a persistent $\text{sech}$-shaped profile.  In some cases, initial $\text{sech}$-shaped profiles can become multi-humped or asymmetric, violating our assumed profile shape.  However, over short timescales the $\text{sech}$ profile is nearly preserved.  We mention that these ``short'' timescales correspond to physical time scales of a few minutes in a field with a spatial wavelength of 200 m (typical in the deep ocean).

We intend to apply this methodology to wave group evolution in the 2D MNLS equation and compare these dynamics with existing results for 2D NLS \cite{adcock2012}. Moreover, this analysis will be fruitful in developing a scheme to predict extreme waves before they occur--we can use this analysis to determine which groups in an irregular field are likely to focus and create an extreme wave \cite{cousinsSapsis2015_PRSA}. We alse plan to use this approach to develop quantification schemes for the heavy tailed statistics of such intermittently unstable systems using a total probability decomposition \cite{mohamad2015_JUQ}. Finally, this adaptive projection approach respecting invariant properties of the solution introduces a new paradigm that will be useful for other systems involving energy localization \cite{chung2011,erkintalo2011}.    

The authors greatfully acknowledge support by the Naval Engineering Education Center (NEEC) grant 3002883706 and ONR grant N00014-14-1-0520.

\bibliographystyle{apsrev4-1}
\bibliography{waveRefs}

\end{document}